# End-to-end 5G services via an SDN/NFV-based multi-tenant network and cloud testbed

Raul Muñoz, Josep Mangues-Bafalluy, Nikolaos Bartzoudis, Ricard Vilalta, Ricardo Martínez, Ramon Casellas,
Nicola Baldo, José Núñez-Martínez, Manuel Requena-Esteso, Oriol Font-Bach, Marco Miozzo, Pol Henerejos,
Ana Pérez-Neira, Miquel Payaró
Centre Tecnològic de Telecomunicacions de Catalunya (CTTC)
Castelldefels (Barcelona), Spain
name.lastname@cttc.es

*Abstract*—5G has a main requirement of highly flexible, ultra-low latency and ultra-high bandwidth virtualized infrastructure in order to deliver end-to-end services. This requirement can be met by efficiently integrating all network segments (radio access, aggregation and core) with heterogeneous wireless and optical technologies (5G, mmWave, LTE/LTE-A, Wi-Fi, Ethernet, MPLS, WDM, software-defined optical transmission, etc.), and massive computing and storage cloud services (offered in edge/core data centers). This paper introduces the preliminary architecture aiming at integrating three consolidated and standalone experimental infrastructures at CTTC, in order to deploy the required end-to-end top-to-bottom converged infrastructure pointed out above for testing and developing advanced 5G services.

*Keywords—5G testbeds, Software Defined Networks, Network Function Virtualization, Network Function Split.*

## I. INTRODUCTION

In order to build the end-to-end network testbed, CTTC will leverage on its already existing experimental facilities. The existing experimental facilities cover complementary technologies from terminals to radio access, aggregation/core and cloud and are, namely: the GEDOMIS® testbed (LTE/5G PHY testbed) [1], the EXTREME Testbed® (wireless HetNet and backhaul, and edge data-center) [2], and the ADRENALINE Testbed® (packet aggregation and optical core network, core data-center) [3], as shown in Fig. 1. In the following sections, we present two uses cases addressing Fixed Mobile Convergence (FMC) developed in ADRENALINE and EXTREME, and virtual mobile network function splitting and deployment, involving EXTREME and GEDOMIS.

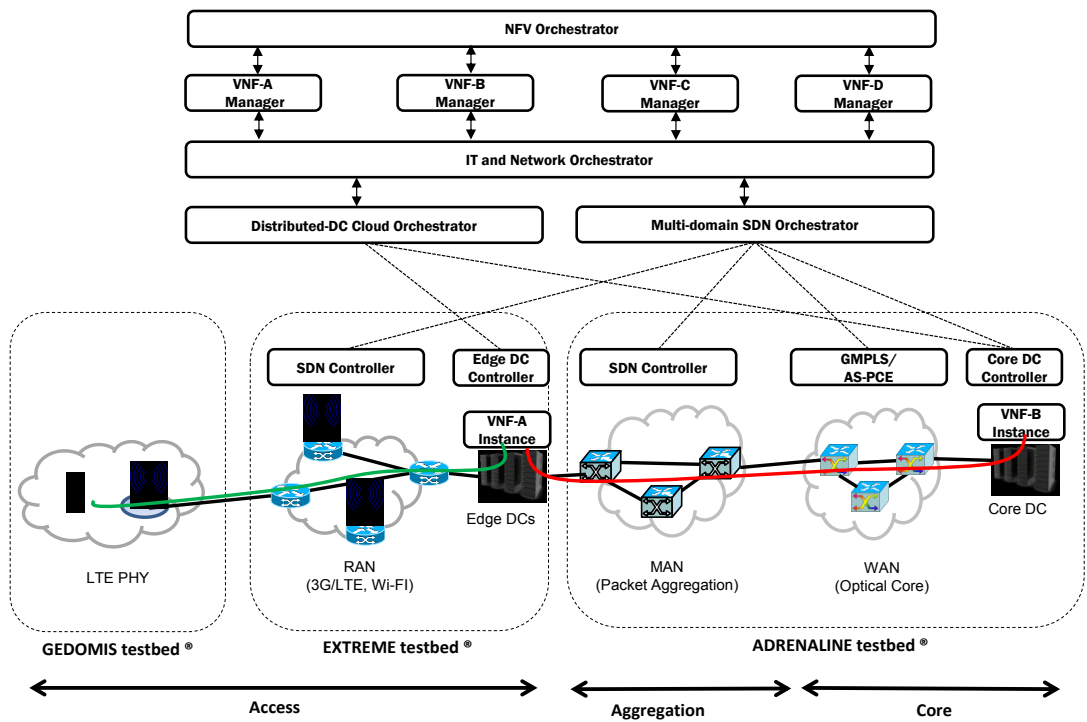

Fig. 1. CTTC SDN/NFV-based multi-tenant network and cloud testbed for end-to-end 5G services.

## II. SDN/NFV Network and Cloud Computing Platform Architecture

Software Defined Networking (SDN) [4] and Network Functions Virtualization (NFV) [5] technologies are the key enablers to federate heterogeneous experimental facilities and to integrate both network and cloud resources to offer advanced end-to-end 5G services upon multi-domain heterogeneous networks and distributed data centers (DC). SDN has emerged as the most promising candidate to improve network programmability and dynamic adjustment of the network resources. SDN is defined as a control framework that supports the programmability of network functions and protocols by decoupling the data plane and the control plane, which are currently integrated vertically in most network equipment. SDN proposes a logically centralized architecture where the control entity (SDN controller) is responsible for providing an abstraction of network resources through Application Programming Interfaces (API). One of the main benefits of this architecture resides on the ability to perform control and management tasks of different wireless and wired network forwarding technologies (e.g., packet/flow switching or circuit switching) by means of the same network controller. The OpenFlow protocol is the most commonly deployed protocol for enabling SDN. It offers a logical switch abstraction, mapping high-level instructions of the protocol to hide vendor-specific hardware details, which mitigates inter-operability issues commonly found in multi-vendor deployments [6]. This abstraction enables SDN to perform network virtualization, that is, to slice the physical infrastructure and create multiple co-existing network slices (virtual networks) independent of the underlying wireless or optical technology and network protocols. In a multi-tenant environment, these virtual networks can be independently controlled by their own instance of SDN control plane (e.g., virtual operators) [7].

Ideally, the SDN architecture is based on a single control domain comprising multiple network nodes featuring diverse technologies provided by different vendors that are controlled through standard interfaces. However, network operators usually fragment their networks into multiple domains to cope with administrative and regional organizations. Each domain can be provided by a different vendor with its own control plane technology (SDN/OpenFlow with some proprietary extensions, legacy GMPLS/PCE, MPLS control, etc.). In our 5G experimental platform, the RAN provided by EXTREME and GEDOMIS is controlled by an SDN controller, and the MAN and WAN provided by ADRENALINE are controlled by an SDN controller and a distributed GMPLS control plane with an Active Stateful PCE, respectively. Thus, a multi-domain network orchestration mechanism is required, as proposed in Fig. 1. The network orchestration mechanism acts as unified network operating system (or controller of controllers) allowing the composition, at a higher, abstracted level, of end-to-end virtual infrastructures [8] as well as end-to-end provisioning services [9] across multiple domains with heterogeneous control and network technologies. SDN is a suitable candidate for end-to-end network service orchestration due to its centralized control nature and the standard and open northbound APIs.

5G will enable the collection of a huge amount of data generated at the terminals, sensors, machines, nodes, etc., that will be transported through networks to data-centers in order to be processed (Big Data) and make the proper decisions (Cognition). The notion of NFV relates to deploying network functions that are typically deployed in specialized and dedicated hardware servers, as software instances (named virtual network functions – VNF) running on commodity servers in data-centers (DCs) through software virtualization techniques. The rise of NFV will also require investments in cloud/DC [10]. Examples of VNFs include IP network functions such as load balancers, firewalls, security or Authentication, Authorization and Accounting (AAA), LTE/EPC network functions, such as Mobility Management Entity (MME), Serving Gateway (SGW), and PDN Gateway (PGW), or transport network functions [11][12]. Originally, cloud computing services have been offered in centralized DCs. However, there is a general trend to spread the DCs to the edge of the network in order to reduce services' latency to the end user. The extension of cloud computing and services to the edge of the network is known as fog computing, and it will lead to an exponential growth on the inter-datacenter traffic requiring high-bandwidth and low-latency 5G networks to interconnect them and offer global end-to-end cloud services. In our 5G end-to-end experimental platform shown in Fig.1, EXTREME contributes with edge DCs close to the end users connected to the RAN, and ADRENALINE with core DCs connected the optical core network. In our approach, global virtualization of IT resources in distributed DCs is provided by means of a Distributed-DC Cloud Orchestrator, responsible of logically managing multiple distributed Cloud Controllers in a federation of multi-cloud testbeds [13], where each Cloud Controller provides Infrastructure as a Service and may be deployed with different cloud computing software platforms, such as OpenStack, CloudStack or OpenNebula. The Cloud Orchestrator takes over the creation/migration/deletion of VM instances (computing service), disk images storage (image service), and the management of the VM network interfaces (networking service) from any DC.

The interconnection of different DC sites that are physically dispersed, but logically centralized is one of the major challenges to face in order to provide global end-to-end cloud and NFV services. VNFs running on top of a VM can be located in the most appropriate DC, and can be interconnected between them and with the end users (i.e., terminals) in a certain way (forwarding graph) in order to achieve the desired overall end-to-end functionality or service. This is known as "service chaining". Thus, VNFs can be distributed over several DCs connected through multiple heterogeneous wireless and optical networks. Consequently, there is the need to perform an integrated orchestration of cloud (IT) and network resources to dynamically deploy virtual machines and provide the required network connectivity between DCs and between DCs and end-users [14]. In our architecture, this function is provided by the IT and Network Orchestrator, as shown in Fig. 1, and is responsible of effectively coordinating the management of the IT resources in the distributed DCs, and the network resources in the heterogeneous optical and wireless networks, providing a unified operating system towards the applications, such as the VNF managers. The VNF manager is responsible for the lifecycle management (i.e., creation, configuration, and removal) of a Virtual Network Function. Finally, the NFV Orchestrator is the responsible for managing the life cycle of the different VNFs and the required network resources in order to deploy end-to-end NFV forwarding graphs.

### III. USE CASE I: FIXED MOBILE CONVERGENCE

This use case [15] investigates a unified access and aggregation network architecture allowing fixed and mobile networks to converge (Fixed Mobile Convergence, FMC). This convergence of fixed and mobile networks will be driven by an improved network infrastructure ensuring reduced cost (both OpEx and CapEx). The new FMC structure will thereby inherently improve end user's quality of experience, e.g., due to reduced delay, increased throughput and seamless access to broadband networks. Convergence of key fixed and mobile network functions will also improve ease of operation for network operators.

One of the approaches addressed is the centralized functional convergence solution depicted in Fig.2. The goal is to deploy a common and unified orchestration system based on the SDN principles to seamlessly handle the automatic provisioning and recovery of both fixed and mobile data flows. In the example, a multi-layer aggregation infrastructure provided by the ADRENALINE testbed is considered. This network combines the benefits of both statistical multiplexing of packet switching (MPLS-TP) and huge transport capacity of optical switching (Flexi-grid DWDM). Thereby, both fixed and mobile services are grouped at the packet layer and transported over the same optical tunnels. For example, mobile bearers are encapsulated at the MPLS layer and then transported transparently through the backhaul, i.e., the 3GPP communication between mobile base stations (eNodeBs) and the Evolved Packet Core (EPC). The mobile network (RAN and EPC building blocks and protocol stacks) is provided by the LENA emulator [16] running over the EXTREME testbed.

The overall goal of this use case is the harmonic integration and automation of the control planes of the mobile network layer and the transport network layer, which have hitherto operated quite independently. SDN offers an appropriate framework in this direction. In this sense, Fig.2 shows how the mobile network layer entities, when establishing a bearer, for instance, and by appropriately interacting with the transport network controller, automatically trigger the establishment of a transport path able to fulfil the QoS requirements of that specific bearer.

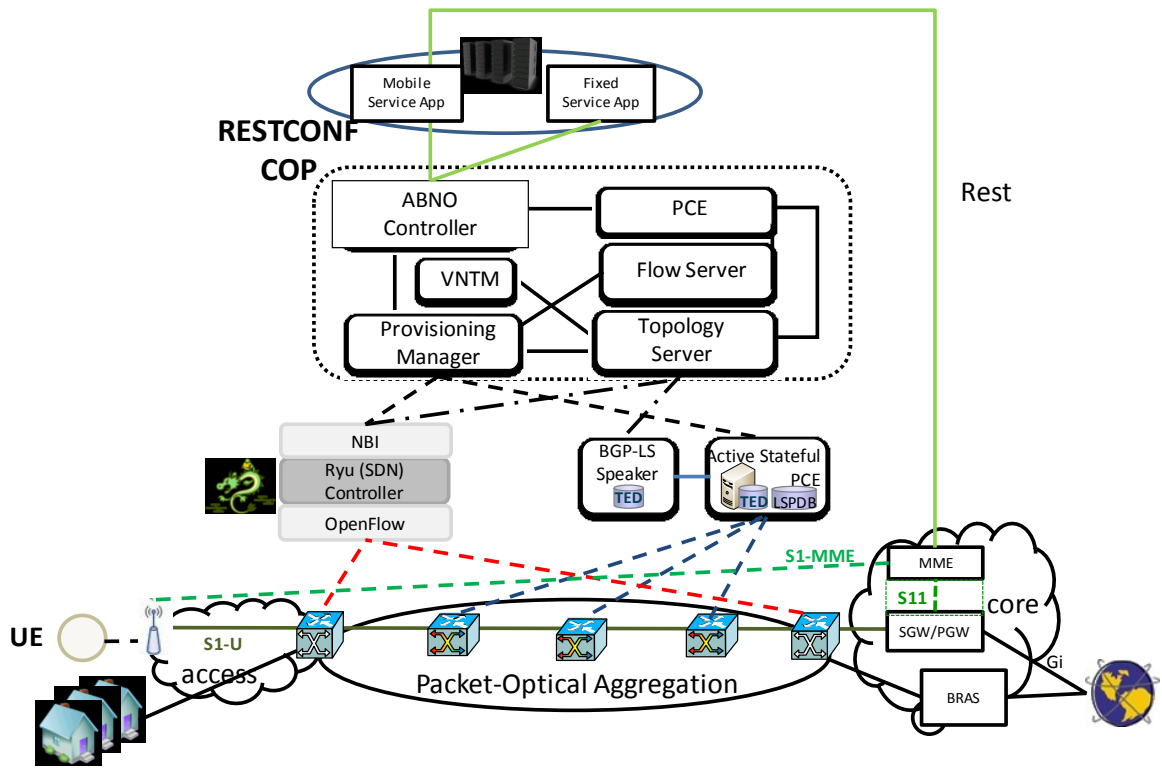

Fig. 2. Fixed Mobile Convergence use case.

The control of the packet and optical network elements is provided by two independent control planes, which are coordinated by a higher-level orchestration system. For packet switching, an SDN control plane is used, which commands the forwarding configuration via the OpenFlow protocol. On the other hand, the optical infrastructure (WSON) is governed by an Active Stateful PCE, which is the responsible for computing and triggering the optical LSP establishment via the GMPLS signaling. The role of the SDN-based orchestration (based on the IETF Application-Based Network Operations, ABNO) system is twofold: first, it allows the coordination of the heterogeneous control plane instances for governing the multi-layer aggregation network; second, it serves the requests for both fixed and mobile connections arriving from the respective service applications located in the cloud. Notice that such service applications running on top of the orchestrator are responsible for requesting, modifying and releasing the connections to serve both fixed and mobile services.

IV. USE CASE II: VIRTUAL MOBILE NETWORK FUNCTION SPLITTING AND DEPLOYMENT

In this use case, our goal is not only to integrate LENA running over EXTREME towards the backhaul and core, as explained above, but also with a real-time LTE/5G PHY testbed (GEDOMIS) [17]. In fact, this is one of the key components of our SDN/NFV testbed, as it will substantially widen the scope of the NFV use cases that can be evaluated, e.g., the virtual base station one listed in ETSI NFV ISG use cases document.

Having a complete mobile protocol stack implementation covering from L1 (PHY) to L7 (application) entirely developed by CTTC will open up a whole range of research opportunities for CTTC. For instance:

- Studying the most appropriate hardware-software split of baseband processing functionality depending on the network conditions that account instantaneous performance and energy consumption requirements or are bound to other functional criterions coming from the network operator side.

- Studying the tradeoff in different virtualization strategies for the L2-L4 stack, for example developing proof-of-concepts of virtual base station deployment as a function of network conditions similarly as in the previous case.

- In general, designing and rapidly prototyping candidate 5G technologies at the different layers, especially L1-L4, and evaluating their performance covering layer-specific KPIs as well as end-to-end QoE.

More specifically, the testbed features a complete RAN+EPC testbed, including the FPGA-based LTE/5G PHY provided by the GEDOMIS testbed, the LTE Radio Protocol Stack (MAC, RLC, PDCP, RRC) provided by LENA, and the EPC Protocol Stack (MME, SGW and PGW network entities together with S1, S11 and X2 interfaces) also from LENA. All these components have been developed at CTTC, hence the resulting testbed would have a great flexibility for CTTC researchers and be much preferable as a research platform than closed commercial equipment. Fig. 3 describes the integration in terms of protocol stack (left) and one possible functional split (right).

Overall, the EXTREME-GEDOMIS integration completes the end-to-end 5G CTTC testbed by providing full control of the communication path at all layers of the protocol stack, for all network planes (data, control, and management), through all network segments (access, aggregation, and core), through wireless and optical networks. We believe this represents a powerful experimentation framework for the evaluation of any relevant 5G use case

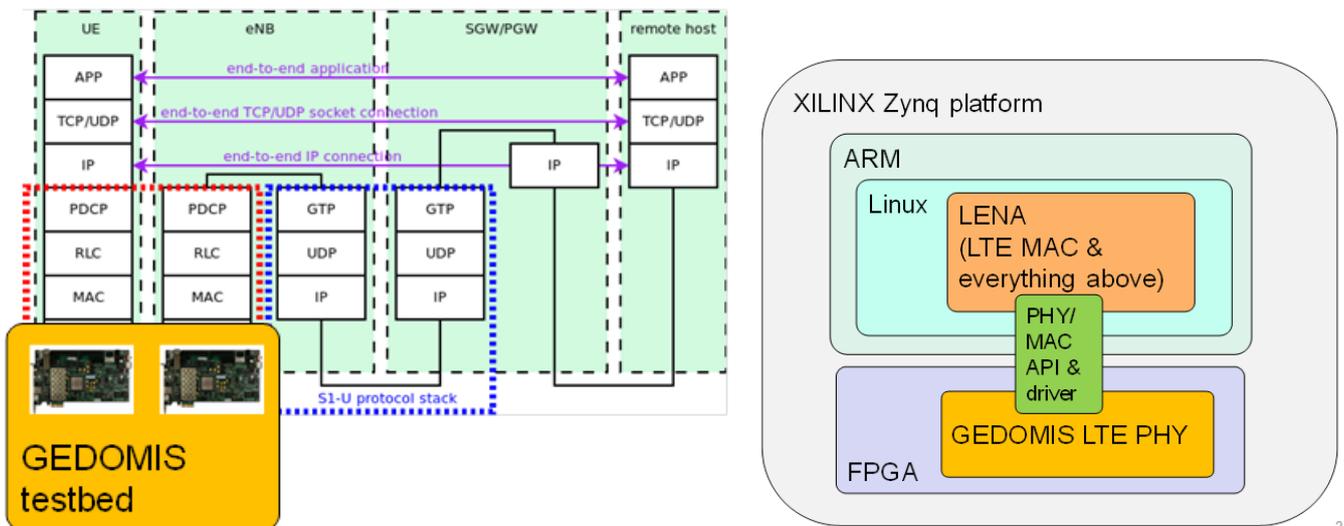

Fig. 3. Virtual mobile network function splitting and deployment use case.

## V. CONCLUSIONS

This paper has presented the preliminary architecture of an SDN/NFV network and cloud computing platform for end-to-end 5G services. This platform integrates ADRENALINE, GEDOMIS and EXTREME testbeds, three complementary testbeds developed by CTTC, spanning from terminals to radio access network, aggregation/core networks and cloud. One of the challenges to face is to deploy a multi-domain network orchestration mechanism to offer dynamic and flexible end-to-end connectivity and virtual network provisioning services across multi-domain and multi-technology networks, integrating all network segments (radio access, aggregation and core) with heterogeneous wireless and optical technologies. The second main challenge is to perform an integrated orchestration of distributed cloud resources (virtual compute and storage) and network resources to dynamically deploy virtual machines or VNF instances (including RAN, transport, and mobile core functions) and provide the required network connectivity between DCs and between DCs and end-users. Two preliminary integration scenarios have been defined. The former integrates ADRENALINE and EXTREME to deploy a Fixed Mobile Convergence use case. The latter integrates EXTREME and GEDOMIS to deploy a use case on virtual mobile network function splitting and deployment.

ACKNOWLEDGEMENTS

The GEDOMIS®️ testbed, the EXTREME®️ testbed, and the ADRENALINE®️ Testbed, are partially financed by the Operational European Regional Development Fund Programme Catalonia 2007-2013. This work has been supported by the Catalan Government under grants 2014 SGR 1397, 2014 SGR 1567, 2014 SGR 1551 and by the EC under project Network of Excellence in Wireless Communications (Newcom#, Grant Agreement 318306).